\documentstyle[pra,aps,twocolumn,epsf,amssymb]{revtex}

\begin{document}
\draft
\title{Comment on ``Dispersion-Independent High-Visibility Quantum
Interference $\ldots$"}
\author{Yoon-Ho Kim, Sergei P. Kulik,\thanks{Permanent address:
Department of Physics, Moscow State University, Moscow, 119899,
Russia.} Morton H. Rubin, and Yanhua Shih}
\address{Department of Physics, University of Maryland, Baltimore
County, Baltimore, Maryland 21250}
\date{Revised 05 September, 2000}

\maketitle

\vspace*{-10mm}Recently Atat\"{u}re {\em et al.} claimed to
``recover" high-visibility quantum interference in femtosecond
pulse pumped type-II Spontaneous Parametric Down-Conversion
(SPDC) using neither spectral post-selection nor a thin nonlinear
crystal [1]. We show in this Comment that the interpretation of
experimental data as well as the theory presented in Ref.[1] are
incorrect and discuss why such a scheme cannot be used to
``recover" high-visibility quantum interference.

Let us first discuss the theory presented in Ref.[1]. Equation (8)
is incorrect and, consequently, so is Eq.(10). According to
Eq.(10), the coincidence counting rate should have a
$\sin^2(\theta_1+\theta_2)$ modulation with 100\% visibility for
{\em arbitrary angles of $\theta_1$ and $\theta_2$} when
$\tau=0$. As we shall see in our experiment, this is not so. This
is because for arbitrary $\theta_1$ and $\theta_2$, there should
be two more terms,i.e.,
$\cos(\pi/4-\theta_1)\sin(\pi/4-\theta_2)[{\mathcal{A}}(t_1,t_2+\tau)-
{\mathcal{A}}(t_2+\tau,t_1)]-\sin(\pi/4-\theta_1)\cos(\pi/4-\theta_2)[{\mathcal{A}}(t_1+\tau,t_2)-
{\mathcal{A}}(t_2,t_1+\tau)]$, which cannot be ignored in Eq.(8).
These two terms have no overlap if $\tau=0$. This will reduce the
visibility of the polarization correlation at arbitrary
$\theta_1$ and $\theta_2$ except at the $H$ and $V$ settings of
the analyzers.

\begin{figure}[tbp]
\centerline{\epsfxsize=3.4in\epsffile{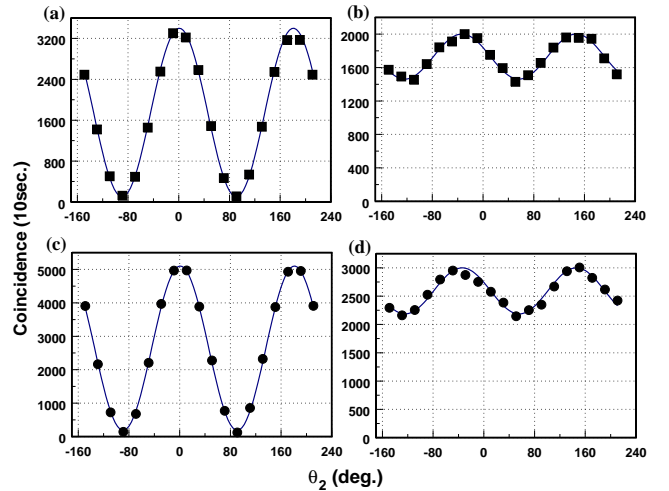}}\caption{Experimental
data. With ``$X$-$Y$ delay"($\tau=0$): (a) $\theta_1=90^\circ$,
(b) $\theta_1=45^\circ$. Without ``$X$-$Y$ delay": (c)
$\theta_1=90^\circ$, (d) $\theta_1=45^\circ$. }\label{fig:data}
\end{figure}

To demonstrate Eq.(10) in Ref.[1] is incorrect, we have performed
an experiment which is identical to Fig.1 in Ref.[1] in which the
polarization correlation is measured. When $\theta_1=90^\circ
(H)$ or $0^\circ (V)$ high-visibility modulation is observed as
$\theta_2$ is varied, see Fig.\ref{fig:data}(a). This is what
Atat\"{u}re {\em et al.} observed in Ref.[1]. However, at
$\theta_1=45^\circ$, the visibility is immediately reduced to
16\% [Fig.\ref{fig:data}(b)].

This means that the ``$X$-$Y$ delay" at $\tau=0$ does not
``recover" the quantum interference as the authors expected. In
fact, one can observe the same interference pattern when the
``$X$-$Y$ delay" is absent. To show this, we removed the
``$X$-$Y$ delay" from the setup, set $\theta_1=90^\circ$, and
varied $\theta_2$. The ``visibility" is $\approx$100\%, see
Fig.\ref{fig:data}(c).  Setting $\theta_1=45^\circ(H)$ and varying
$\theta_2$ again, as evident from Fig.\ref{fig:data}(d), the
visibility is as low as 16\%. This demonstrates that the ``$X$-$Y$
delay" has no net physical effect when $\tau=0$. This also shows
that what is observed in Ref.[1] is not quantum interference. It
simply shows that the signal is $V$-polarized and the idler is
$H$-polarized.

These data clear show that $|V\rangle|H\rangle$ has not been
transformed to $|X\rangle|X\rangle-|Y\rangle|Y\rangle$ as the
authors claim [Eq.(10)]. In fact, such a ``cascaded
transformation of the two-photon state" cannot occur unless
proper longitudinal \newpage \vspace*{-10mm} \noindent
compensation is made first [2]. Therefore, it is obvious that
this kind of scheme cannot be used to ``recover" quantum
interference. We also note that Fig.3 in Ref.[1] might lead to
confusion since readers might mistakenly consider it to show
space-time interference. In fact, only polarization correlation
measurement is observed in Ref.[1] at a fixed angle
$\theta_1=0^\circ$.

It is true that Atat\"{u}re {\em et al.} made some kind of
polarization state transformation of biphotons. Certainly these
transformations are related to $\tau$ and the pump pulse duration
(For a general description of polarization transformation of
biphotons, see Ref.[3]). It, however, has nothing to do with the
``recovery" of quantum interference as they claim.

In conclusion, we have experimentally and theoretically shown that
Atat\"{u}re {\em et al.}'s claim to be in error. Neither the
experimental data nor the correct theory support their claim.
Finally, we would like to mention that we have recently developed
a new method of generating entangled photon pairs pumped by
femtosecond pulses which shows true high-visibility quantum
interference [4]. \\

\noindent$[1]$ M. Atat\"{u}re, A.V. Sergienko, B.E.A. Saleh, and
M.C. Teich, Phys. Rev. Lett. {\bf 84}, 618 (2000). \\
$[2]$ M.H. Rubin, D.N. Klyshko, Y.H. Shih, and A.V. Sergienko,
Phys. Rev. A {\bf 50}, 5122 (1994).\\
$[3]$ A.V. Burlakov and D.N. Klyshko, JETP Lett. {\bf 69}, 839
(1999).\\$[4]$ Y.-H. Kim, S.P. Kulik, and Y.H. Shih, Phys. Rev. A.
{\bf 62}, 011802(R) (2000); \verb+quant-ph/0007067+.\\

\end{document}